\documentclass[twocolumn,showpacs,preprintnumbers,amsmath,amssymb,floatfix]{revtex4}

\usepackage[normalem]{ulem}
\usepackage{verbatim}
\usepackage{graphicx}
\usepackage{dcolumn}
\usepackage{bm}
\usepackage{xcolor}
\usepackage{url}
\usepackage{amsmath}
\usepackage{morefloats}
\usepackage{times}
\usepackage[varg]{txfonts}
\usepackage{multirow}
\usepackage{lineno}
\usepackage{hyperref}
\usepackage{notoccite}

\usepackage{color}

\graphicspath{{./figures/}}
\newcommand{\be}{\begin{equation}}
\newcommand{\ee}{\end{equation}}
\newcommand{\ba}{\begin{eqnarray}}
\newcommand{\ea}{\end{eqnarray}}
\newcommand{\nn}{\nonumber}

\def\ltsima{$\; \buildrel < \over \sim \;$}
\def\simlt{\lower.5ex\hbox{\ltsima}}
\def\gtsima{$\; \buildrel > \over \sim \;$}
\def\simgt{\lower.5ex\hbox{\gtsima}}
\begin{document}

\title[blah]{A morphology-independent data analysis method for detecting and characterizing 
gravitational wave echoes}

\author{Ka Wa Tsang$^{1}$, Michiel Rollier$^{1}$, Archisman Ghosh$^{1}$, 
Anuradha Samajdar$^{1}$, Michalis Agathos$^{2}$, Katerina Chatziioannou$^{3}$, Vitor Cardoso$^{4}$, Gaurav Khanna$^{5}$, and Chris Van Den Broeck$^{1,6}$}

\affiliation{$^1$Nikhef -- National Institute for Subatomic Physics, 105 Science Park, 
1098 XG Amsterdam, The Netherlands \\
$^2$DAMTP, Centre for Mathematical Sciences, University of Cambridge, Wilberforce Road, 
Cambridge CB3 0WA, United Kingdom \\
$^3$Canadian Institute for Theoretical Astrophysics, 60 St.~George Street, University of Toronto, 
Toronto, ON M5S 3H8, Canada \\
$^4$CENTRA, Departamento de F\'{i}sica, Instituto Superior T\'{e}cnico -- IST,
Universidade de Lisboa -- UL, Avenida Rovisco Pais 1, 1049 Lisboa, Portugal \\
$^5$Department of Physics and Center for Scientific Computing and Visualization Research, 
University of Massachusetts Dartmouth, North Dartmouth, MA 02747, USA \\
$^6$Van Swinderen Institute for Particle Physics and Gravity, University of Groningen, \\
Nijenborgh 4, 9747 AG Groningen, The Netherlands}

\date{\today}

\begin{abstract}
The ability to directly detect gravitational waves has enabled us to empirically probe the nature 
of ultra-compact relativistic objects. Several alternatives to the black holes of classical 
general relativity have been proposed which do not have a horizon, in which case a newly 
formed object (\emph{e.g.}~as a result of binary merger) may emit \emph{echoes}: bursts of gravitational 
radiation with varying amplitude and duration, but arriving at regular time intervals. Unlike in 
previous template-based approaches, we present a morphology-independent search method to find echoes in the data from 
gravitational wave detectors, based on a decomposition of the signal in terms of generalized wavelets 
consisting of multiple sine-Gaussians. The ability of the method to discriminate between echoes and
instrumental noise is assessed by inserting into the noise two different signals: a train of sine-Gaussians, 
and an echoing signal from an extreme mass-ratio inspiral
of a particle into a Schwarzschild vacuum spacetime, with reflective boundary conditions close to the horizon. 
We find that 
both types of signals are detectable for plausible signal-to-noise ratios in existing detectors
and their near-future upgrades. Finally, we show how the algorithm can provide a characterization 
of the echoes in terms of the time between successive bursts, and damping 
and widening from one echo to the next. 
\end{abstract}

\pacs{04.40.Dg,04.70.Dy,04.80.Cc}

\date{\today}

\maketitle

\emph{Introduction.} Since 2015, the twin Advanced LIGO observatories \cite{TheLIGOScientific:2014jea} have regularly detected 
gravitational wave (GW) signals from coalescing compact binary objects 
\cite{Abbott:2017xlt,Abbott:2016nmj,TheLIGOScientific:2016pea,Abbott:2017vtc,Abbott:2017gyy}. 
Recently Advanced Virgo \cite{TheVirgo:2014hva} also joined 
the global network of detectors, leading to further detections, including a binary neutron star merger 
\cite{Abbott:2017oio,TheLIGOScientific:2017qsa}. 
These observations have enabled far-reaching tests of general relativity: for the first time the 
genuinely strong-field dynamics of the theory could be empirically investigated, including the behavior 
of pure vacuum spacetime; and the propagation of gravitational 
waves over large distances 
could be studied, leading to stringent bounds on the mass of the graviton and on violations 
of local Lorentz invariance \cite{TheLIGOScientific:2016src,TheLIGOScientific:2016pea,Abbott:2017vtc}. 
A natural next step is to probe the nature of the compact objects 
themselves. For the more massive compact binary coalescences that were observed, how certain can we be 
that these involved the black holes of classical general relativity? In quantum gravity, Hawking's 
information paradox has led to the suggestion of Planck-scale modifications of black hole horizons 
(firewalls \cite{Almheiri:2012rt}) and other alterations of black hole structure (fuzzballs \cite{Lunin:2001jy}). 
In cosmology, dark matter 
particles have been proposed that congregate into star-like objects \cite{Giudice:2016zpa}. Yet another possibility concerns
stars whose interior consists of self-repulsive, de Sitter spacetime, surrounded by a shell
of ordinary matter (gravastars \cite{Mazur:2004fk}). Finally, there is the idea of boson stars, 
macroscopic objects made up of scalar fields \cite{LieblingLR:2012}. 
What these objects have in common is the absence of a horizon, causing ingoing gravitational 
waves (\emph{e.g.}~resulting from merger) to reflect multiple times off effective radial potential barriers, 
with wave packets leaking out to infinity at regular times; these are called \emph{echoes} 
\cite{Cardoso:2016rao,Cardoso:2016oxy,Cardoso:2017cqb}. For 
an exotic object with mass $M$ and a microscopic correction at the horizon scale of size $\ell$, the 
time between echoes tends to be constant, and well approximated by
$\Delta t \simeq n M \log \left(M/\ell\right)$, 
with $n$ a factor of order unity that is determined by the nature of the exotic object (\emph{e.g.}~$n = 8$
for a wormhole, $n = 6$ for a gravastar, and $n = 4$ for an empty shell) \cite{Cardoso:2016oxy}. As an example, taking $M$ to be 
the detector frame mass of the remnant object resulting from the first gravitational wave
detection GW150914 ($M \simeq 65\,M_\odot$) \cite{Abbott:2017xlt,TheLIGOScientific:2016wfe}, setting $n = 4$, and identifying $\ell$ with the Planck length, 
one has $\Delta t \simeq 117$ ms. This is much longer than the duration of the ``ringdown" of the
remnant (about 3 ms), but at the same time sufficiently short that it would be practical to
search for echoes immediately following the main inspiral-merger-ringdown signal.

In \cite{Abedi:2016hgu,Westerweck:2017hus,Maselli:2017tfq}, template-based searches were proposed using a 
heuristic expression for the echo 
waveforms in terms of $\Delta t$ as well as a characteristic frequency, a damping factor, and 
a widening factor between successive echoes. Though expressions 
exist for echo waveforms from selected exotic objects under various assumptions \cite{Cardoso:2016oxy,Cardoso:2017cqb}, 
concrete calculations have so far only been exploratory~\cite{Mark:2017dnq}. Moreover, there may well be other types 
of objects that also cause echoes but have not yet been envisaged. For this reason, it is 
desirable to have a \emph{generic} search for echoes which can capture and characterize a wide 
variety of different 
waveform morphologies. A commonly used method to search for and reconstruct gravitational wave signals 
of \emph{a priori} unknown form is through the \texttt{BayesWave} algorithm 
\cite{Cornish:2014kda,Littenberg:2014oda}.
Here the output of a network of detectors, ${\bf s}$, is written as 
$\mathbf{s} = \mathbf{R} \ast \mathbf{h} + \mathbf{n_g} + \mathbf{g}$, 
where $\mathbf{R}$ is the response of the network to gravitational waves, $\mathbf{h}$ is
the signal, $\mathbf{g}$ denotes instrumental transients or glitches, and $\mathbf{n_g}$
is a stationary Gaussian noise component. The signal model $\mathbf{h}$ and the glitch model $\mathbf{g}$ 
are both characterized as superpositions of appropriate basis functions, and Bayesian evidences 
can be computed for the associated hypotheses. 
From an observational perspective, the defining difference between signals and glitches is that 
the signal is present in the output of all detectors in the network in a coherent way, 
whereas any instrumental glitch will be present in only a single detector's data stream. Thus,  
if a coherent signal is present in the data, then typically a smaller number of basis functions 
will be needed to reconstruct it than 
to reconstruct incoherent glitches, leading to an Occam penalty for the glitch model; at the same time,
the signal is reconstructed with a superposition of the basis functions.

The choice of basis functions to model signals and glitches with is not unique. Due to their 
simplicity, sine-Gaussians were originally employed and they have been shown to lead to 
efficient detection~\cite{2016PhRvD..93b2002K,Littenberg:2015kpb} and 
reconstruction~\cite{Becsy:2016ofp,Chatziioannou:2017ixj} of a wide range of signal morphologies, 
though more options have been explored~\cite{chirplets}.
In this paper and for the study of echoes we propose generalized wavelets which are 
``combs" of sine-Gaussians, 
characterized by
a time separation between the individual sine-Gaussians as well as a fixed phase shift
between them, an amplitude damping factor, and a widening factor. Even though actual echo 
signals are unlikely to resemble any single generalized wavelet and may not even have 
well-defined values for any of the aforementioned quantities, we do expect superpositions 
of generalized wavelets to be able to capture a wide variety of physical echo waveforms.
Moreover, one can assume the distribution of samples over the generalized wavelet parameter space 
to yield basic information about the structure of the echoes signal, which should then be of
help in identifying the nature of the object that is emitting them. 
%\KC{This needs some care here. The wavelets (generalized or not) are not a complete set. 
%This means that the properties of the individual wavelets need not reflect the properties of 
%the full signal. For example, when we want to calculate the frequency of the signal we do *not* 
%look at the central frequency of each wavelet, but rather measure the frequency from the full signal 
%directly. I can probably convince myself that it is ok to look at the time delay or the damping 
%factor, but I would be careful to not over-interpret the wavelet parameters.}

\emph{Description of the method.} As in the standard \texttt{BayesWave} algorithm, given a 
detector $I$ the signal model in the frequency domain takes the form 
\be
(\mathbf{R} \ast \mathbf{h})_I(f) = \left(F_+^I(\theta, \phi, \psi)\,h_+(f) 
+ F_\times^I(\theta, \phi, \psi)\,h_\times(f)\right)\,e^{2\pi i f \Delta t_I(\theta, \phi)},
\ee
where $h_\times = \epsilon h_+ e^{i\pi/2}$, with $\epsilon$ the ellipticity as in \cite{Cornish:2014kda}. 
The sky position $(\theta, \phi)$ and the polarization
angle $\psi$ are consistent across detectors, whose beam pattern functions are denoted by $F_+^I$ and
$F_\times^I$; $\Delta t(\theta, \phi)$ is the delay between the geocentric and detector arrival times. 
$h_+$ is decomposed into a sum of generalized 
wavelets that are
``combs" of $N_G$ sine-Gaussians in the time domain which are functions of 9 parameters:
\begin{align}
&\Psi(A, f_0, t_0, \tau, \phi_0, \Delta t, \Delta \phi, \gamma, w; t) \nn\\
&= \sum_{n=0}^{N_G} \gamma^n A\,\exp\left[-\left(\frac{t-(t_0 + n\Delta t)}{w^n \tau}\right)^2 \right]\nn\\
& \quad \qquad \times \cos(2\pi f_0 (t - (t_0 + n\Delta t)) + \phi_0 + n \Delta\phi).
\label{generalizedwavelets}
\end{align}
Here $A$ is an amplitude, $f_0$ is a central frequency, $t_0$ is the central time of the first echo, 
$\tau$ is a damping time, $\phi_0$ a reference phase, $\Delta t$ is the time between successive
sine-Gaussians, $\Delta \phi$ is a phase difference between them, $\gamma$ is a damping factor between one 
sine-Gaussian and the next, and $w$ is a widening factor. The glitch model also involves a decomposition 
into the generalized wavelets above. The number of wavelets is allowed to vary. Given $N_d$ detectors, a 
signal described by $N$
generalized wavelets requires $9 N + 4$ parameters to be sampled over 
(the 9 intrinsic parameters and 4 extrinsic ones), while glitches described by $N$ generalized 
wavelets involve $9\,N_d N$ parameters. Hence, when $N_d > 1$ and with a signal
present, the signal model will be preferred over the glitch model because it enables a more parsimonious 
description. 
The noise model consists of colored Gaussian noise whose power spectral density is computed 
using a combination of smooth spline curves and a collection of Lorentzians 
to fit sharp spectral features~\cite{Littenberg:2014oda}.

For each of the three hypotheses, the corresponding parameter space is sampled over using a 
Reversible Jump Markov Chain Monte Carlo algorithm, in which 
the number of generalized wavelets is free to vary as in~\cite{Cornish:2014kda}. 
Evidences for the three hypotheses are then estimated by means of thermodynamic integration, giving  
the Bayes factors $B_{S/N}$ and $B_{S/G}$ for the signal versus noise and signal 
verus glitch hypotheses, respectively.
The samples in parameter space that are produced after a 
%\MA{50,000-sample} 
``burn-in" stage allow us to perform model selection and parameter estimation. Finally, a background 
distribution for $B_{S/N}$ and $B_{S/G}$ is constructed by analyzing 
many stretches of detector noise preceding the main signal. 

\begin{figure}[h]
\includegraphics[width=0.9\columnwidth]{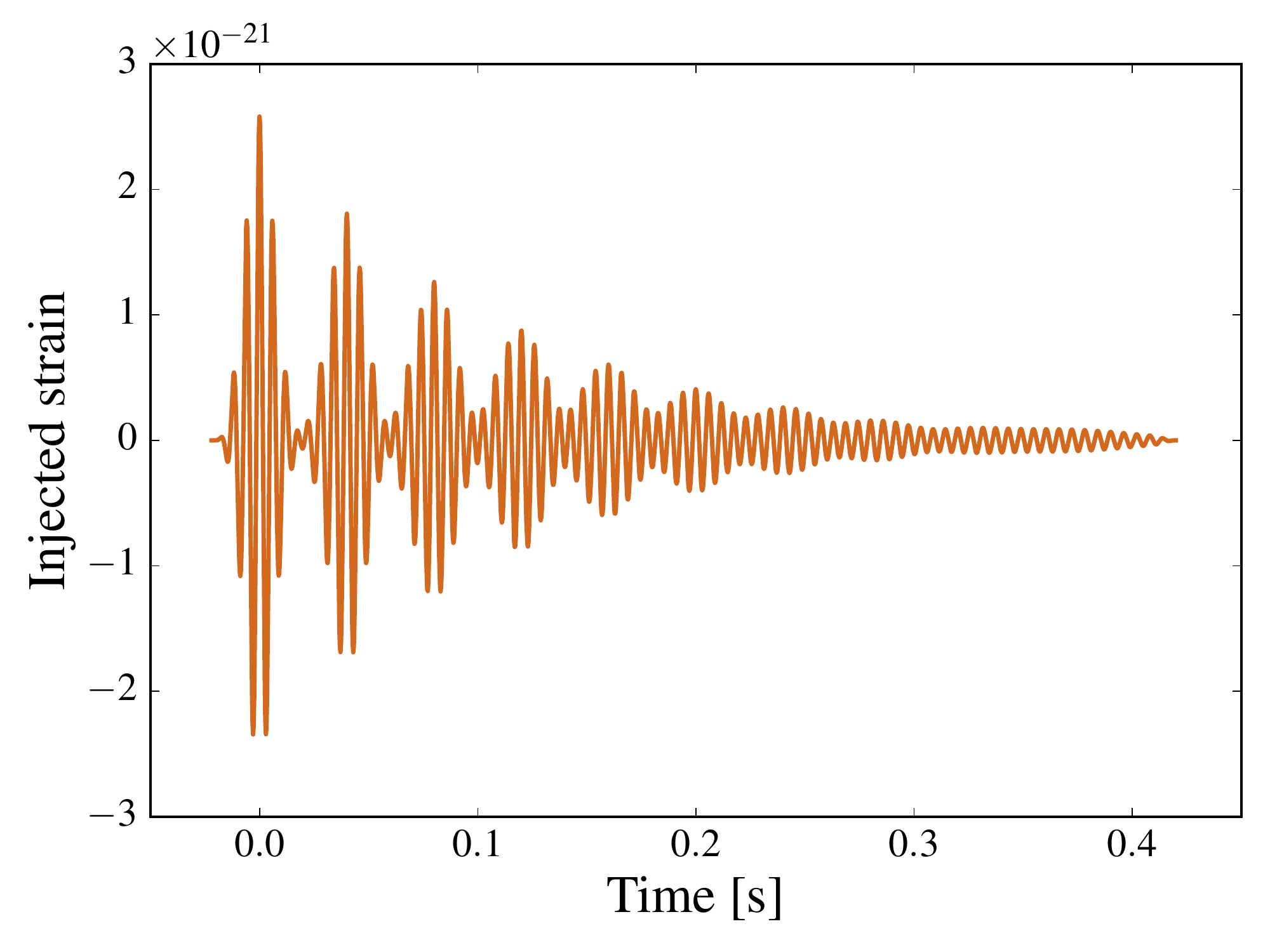}
\includegraphics[width=0.9\columnwidth]{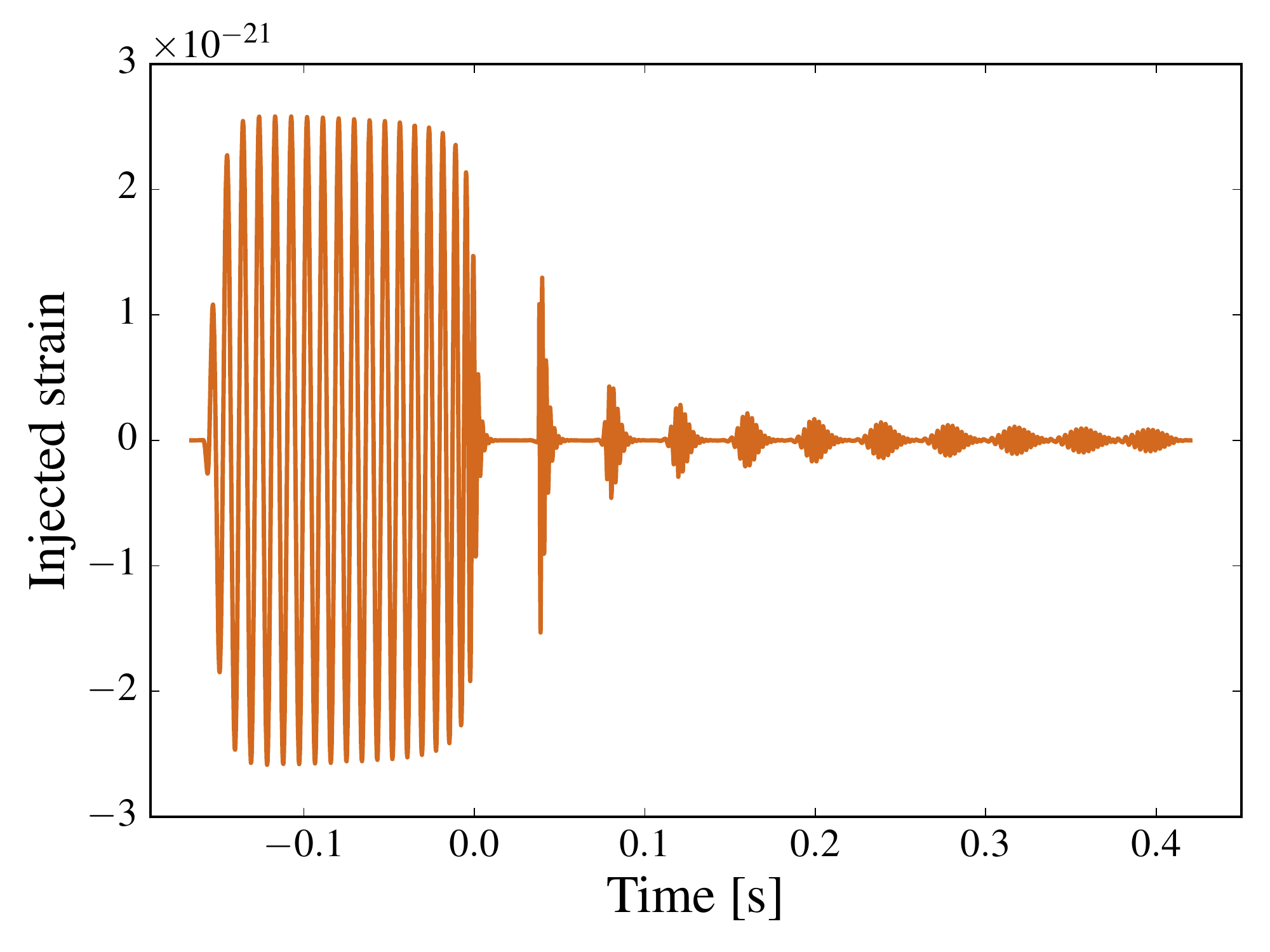}
\caption{The simulated signals used to evaluate the method. Top panel: A train of sine-Gaussians. Bottom panel:
the waveform from a toy model for a mass ratio $q = 1000$ inspiral of a particle in a 
Schwarzschild spacetime, with Neumann reflective boundary conditions just outside the horizon.}
\label{fig:injections}
\end{figure}

\begin{figure}[h]
\includegraphics[width=0.9\columnwidth]{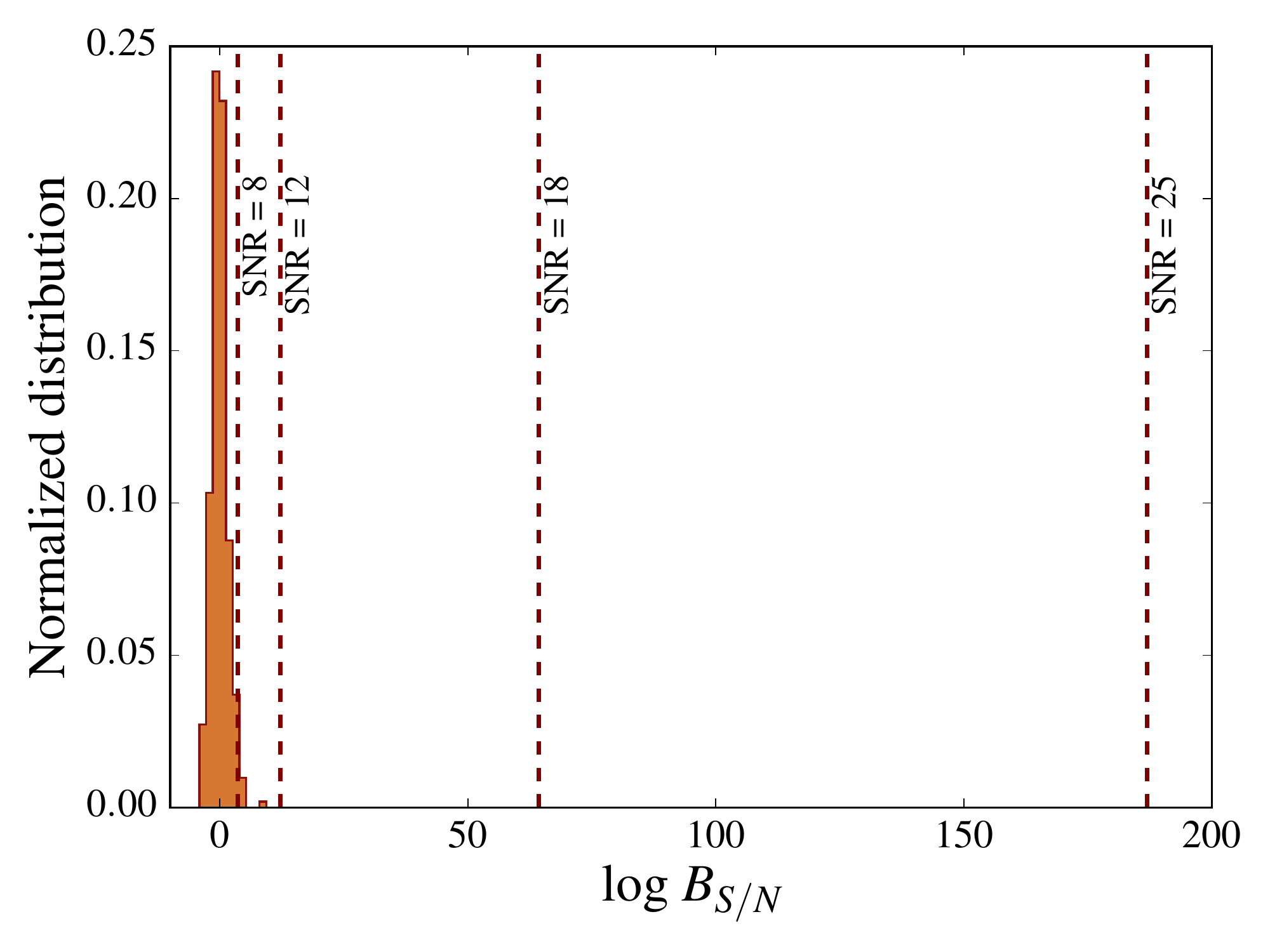}
\includegraphics[width=0.9\columnwidth]{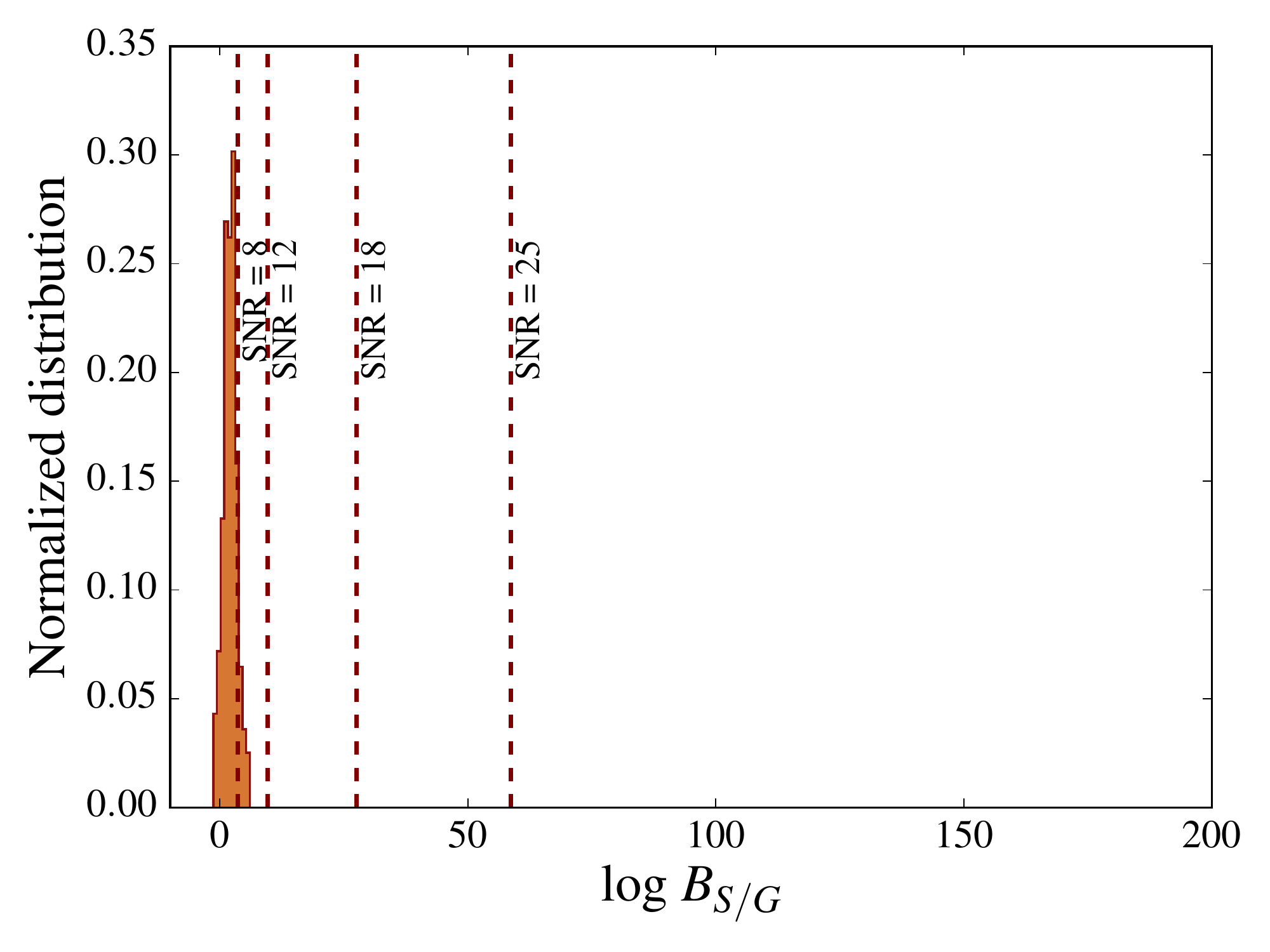}
\caption{Background distributions for the (log) Bayes factors $B_{S/N}$ (top) and $B_{S/G}$ (bottom), 
containing 380 trials. The dashed
lines show the values of these quantities for the injection of echoes from the inspiral toy model with 
SNRs of 8, 12, 18, and 25.}
\label{fig:background}
\end{figure}

\begin{figure}[h]
\includegraphics[width=0.9\columnwidth]{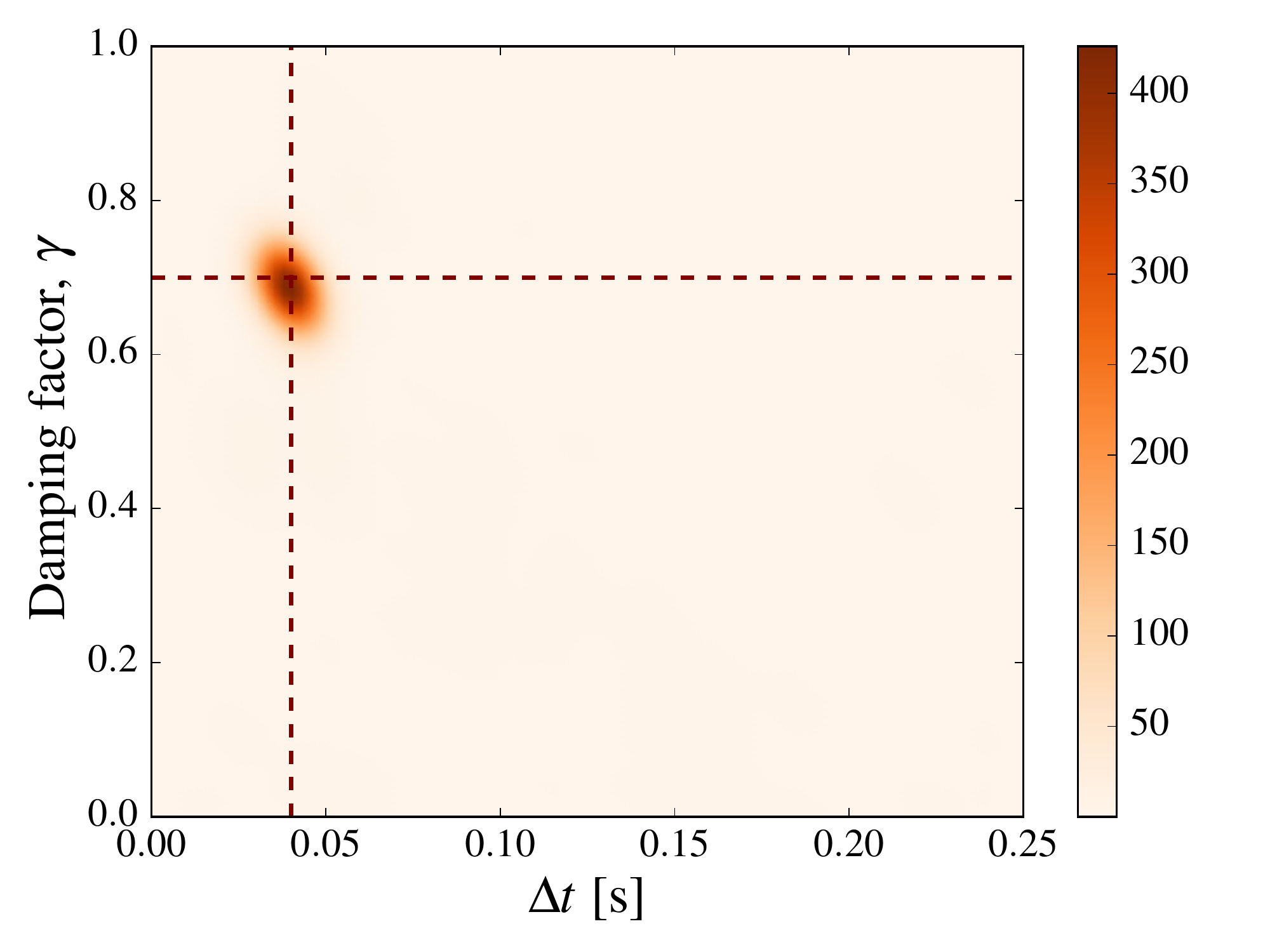}
\includegraphics[width=0.9\columnwidth]{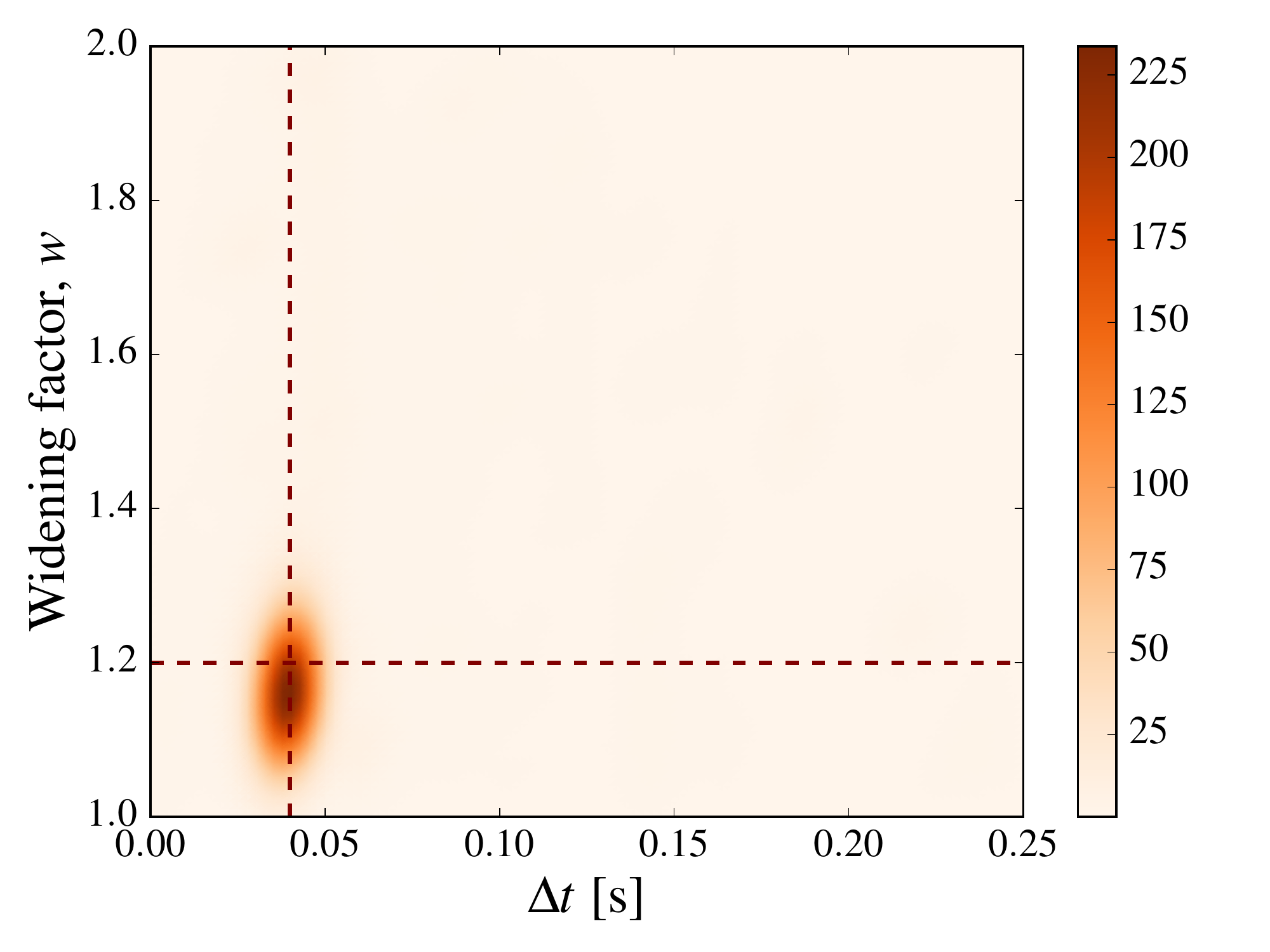}
\caption{The distribution of samples for the case where the injected signal is a comb of sine-Gaussians. 
Top: Damping factor $\gamma$ against the time $\Delta t$ between echoes. Bottom: The widening factor $w$
against $\Delta t$. The colors indicate the number of samples per pixel, 
while the dashed lines show the true values of the parameters.}
\label{fig:pe1}
\end{figure}

\begin{figure}[h]
\includegraphics[width=0.9\columnwidth]{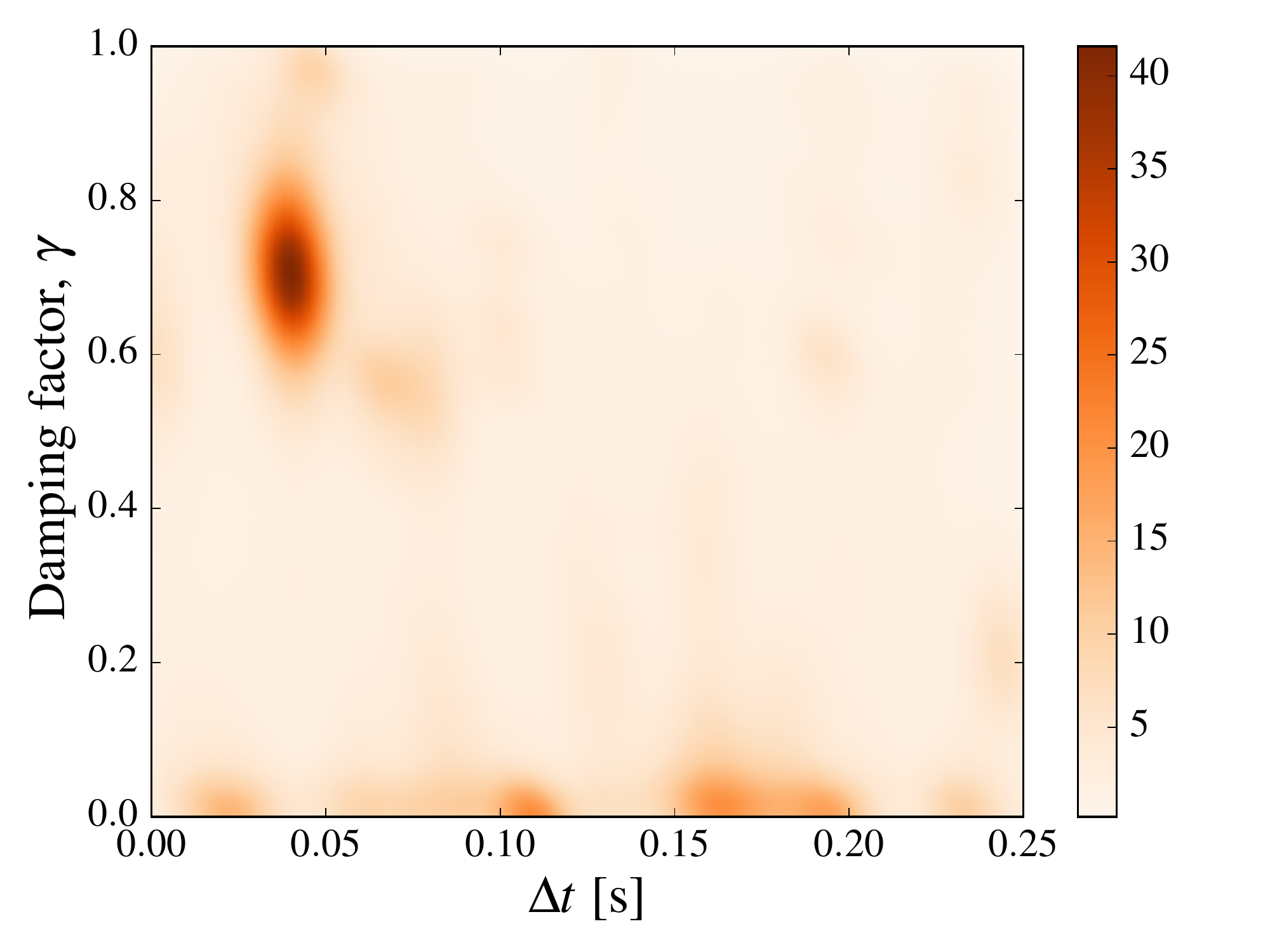}
\includegraphics[width=0.9\columnwidth]{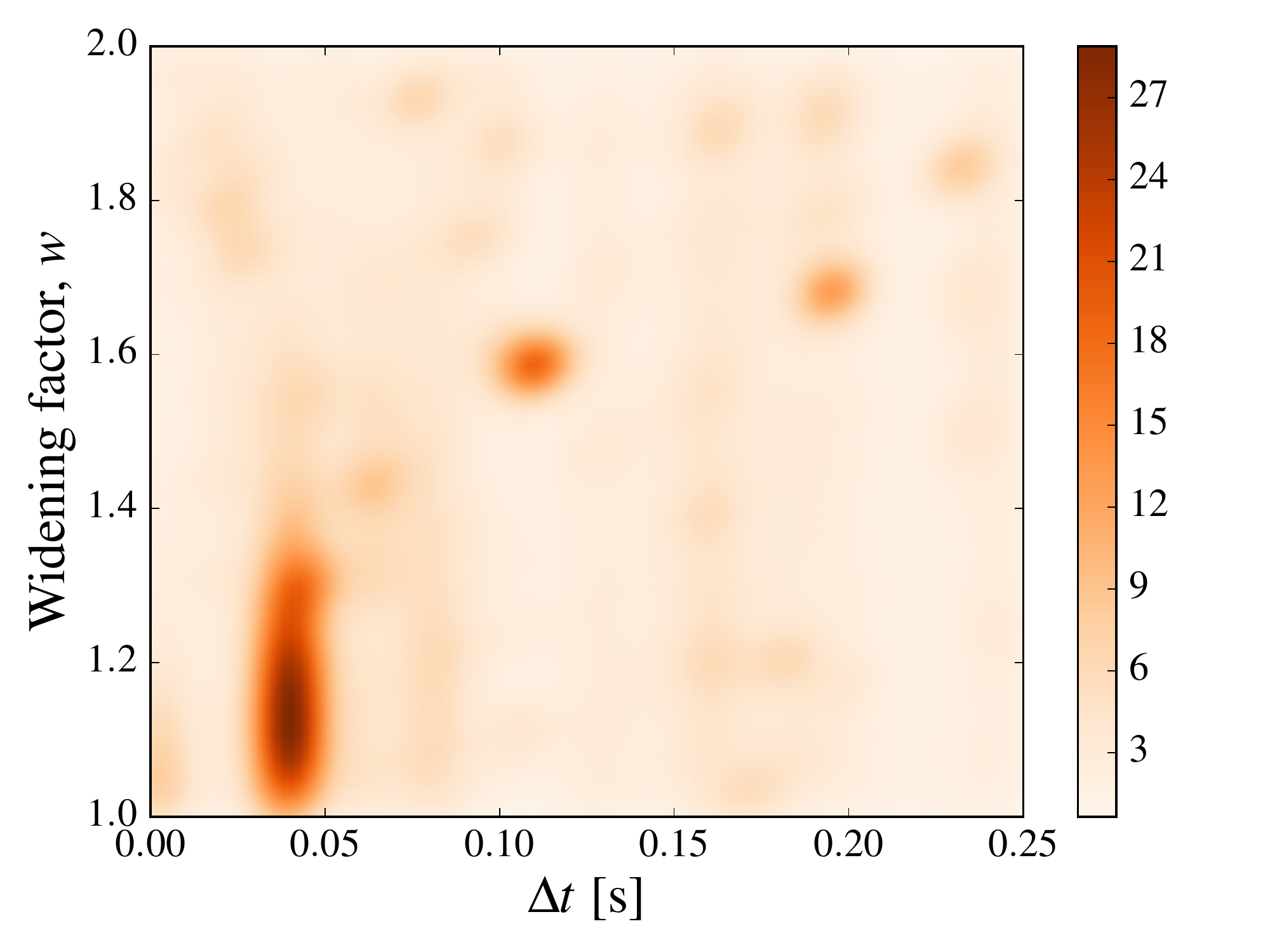}
\caption{The distribution of samples for the case where the injected signal is the inspiral 
toy model. Again we show $\gamma$ versus $\Delta t$ (top) and 
$w$ versus $\Delta t$ (bottom).}
\label{fig:pe2}
\end{figure}

\emph{Results.} In order to test the algorithm we generate stationary, Gaussian noise for a network of two 
Advanced LIGO detectors at 
the predicted
design sensitivity \cite{Aasi:2013wya}. In this we coherently inject (a) a single generalized wavelet as in  
Eq.~(\ref{generalizedwavelets}), and (b) a train of echoes 
from a numerically solved toy model involving the inspiral of a particle in a Schwarzschild spacetime 
with Neumann reflective boundary conditions just outside the horizon, the mass ratio being $q = 1000$
\cite{Price:2017cjr,Khanna:2016yow}. The signals are 
shown in Fig.~\ref{fig:injections}. For case (a), one has $f_0 = 166.7$~Hz, $\tau = 0.0095$ s, 
$\phi_0 = 0$, 
$\Delta t = 0.04$ s, $\gamma = 0.7$, and $w = 1.2$. Both for cases (a) and (b), values for the amplitudes of the injected
signals are chosen such that the combined (matched-filtering) signal-to-noise ratio (SNR) in all echoes is, 
respectively, 8, 12, 18, and 25. The higher values correspond to the SNR in the ringdown signal of a 
gravitational wave detection 
like GW150914 \cite{Abbott:2017xlt} under the assumption that it would be seen in 
Advanced LIGO at final design sensitivity, whereas an SNR of 8 roughly equals the SNR 
that the ringdown actually had for GW150914 \cite{TheLIGOScientific:2016src}. 

For both types of simulated signals, 10 echoes are injected (in reality one would expect infinitely many), 
and the generalized wavelets used to characterize the simulated signals have 5 sine-Gaussians in them.
Case (a) has a well-defined damping factor $\gamma$ and widening
factor $w$, allowing us to establish that the method works as intended, by ascertaining 
that these parameters are recovered 
correctly. In case (b), $\gamma$ and $w$ may not have rigorous meaning, but the 
distributions on parameter space that are obtained
should be indicative of the physics involved; moreover, the peaks of their 
distributions should correspond to what one estimates
from a visual inspection of the signal. In the latter case, 
the stretch of data analyzed excludes the main signal, as one would
also do in reality. In both cases the first echo is searched for in a window for 
$t_0$ that has a width of 0.5 s; for the other parameters the prior distributions are flat in  
$\Delta t \in [0, 0.25]$ s, $\gamma \in [0, 1]$, $w \in [1, 2]$, 
and $\Delta \phi \in [0, 2\pi]$. 

In order to confidently detect echoes, the Bayes factors $B_{S/N}$ and $B_{S/G}$ must be 
compared with a background distribution for these quantities, computed on stretches of detector noise, 
\emph{e.g.}~at times immediately preceding the inspiral-merger-ringdown signal. These are shown in 
Fig.~\ref{fig:background},
together with the values obtained from the injection of echoes for the inspiral toy model. 
For all simulated signals considered here we find that, starting from SNR = 12, $\log B_{S/G}$ and 
$\log B_{S/N}$ are above their respective backgrounds; hence
trains of echoes with this loudness would be detected with confidence. 
It is worth noting that very similar Bayes factors are obtained with 
the original \texttt{BayesWave} algorithm, 
which instead of the generalized wavelets of Eq.~(\ref{generalizedwavelets}) uses the standard Morlet-Gabor 
wavelets consisting of single sine-Gaussians. Hence the use of generalized wavelets does
not significantly improve \emph{detection}. However, the generalized wavelets allow for the 
\emph{characterization} of echoes, to which we now turn. 

Fig.~\ref{fig:pe1} shows the
distribution of samples for case (a), for an SNR of 25 and injected echo-related parameters 
$\Delta t = 0.04$ s, $\gamma = 0.7$, and $w = 1.2$. These are measured correctly, with peak values
and standard deviations $\Delta t = 0.040 \pm 0.007$ s, $\gamma = 0.69 \pm 0.05$, and 
$w = 1.16 \pm 0.09$.  
In Fig.~\ref{fig:pe2} we show the distribution of samples for case (b), again for an SNR of 25; 
visual inspection of the
signal in Fig.~\ref{fig:injections} 
indicates similar values for $\Delta t$, $\gamma$, and $w$
as for case (a), and these are indeed the values where sample distributions have their main peaks.
The peak values and standard deviations are $\Delta t = 0.040 \pm 0.007$~s, $\gamma = 0.71 \pm 0.11$, and 
$w = 1.12 \pm 0.12$. The distribution of $(w, \Delta t)$ samples also shows secondary peaks 
at $3\Delta t$ and $5\Delta t$. These correspond to secondary peaks with $\gamma \simeq 0$ in 
$(\gamma, \Delta t)$ space, which are cases where essentially only one echo was found. However, 
the secondary modes are considerably weaker than the main one.  
Finally, by looking at the injections with SNRs 18, 12, and 8, we checked that measurement 
uncertainties roughly increase with inverse SNR, as expected.
We conclude that, given a sufficiently loud source, not only will we have the ability to detect the presence 
of echoes with high statstical confidence, we will also have a way to infer the properties of the 
exotic compact object.

\emph{Summary.} We have constructed a method to search for, and characterize, gravitational wave echoes in 
a morphology-independent way.
The algorithm decomposes the signal into generalized wavelets taking the form of ``combs" 
of sine-Gaussians in order to capture the essence of echoes in the data. As in the 
original \texttt{BayesWave}, the evidences for three hypotheses 
are compared through a sampling over parameter space: signal, glitch, and Gaussian noise. 
We have shown that for a heuristic but 
physically motivated train of echoes, with plausible loudness given expected detector upgrades, the 
echoes signal can be confidently detected. We expect this to be the case for a wide variety of possible 
signal shapes corresponding to different types of compact objects, irrespective of an object's detailed 
nature; in particular, no template waveforms are needed. Moreover, the distribution of samples over 
parameter space will reveal key characteristics of the echoes such as the time between successive bursts, 
as well as their widening and damping. This information can in turn be used to identify 
the nature of the potentially horizon-less merger remnant. 

\section*{Acknowledgements}

K.W.T., A.G., A.S., and C.V.D.B.~are supported by the 
research programme of the Netherlands Organisation for Scientific Research (NWO). 
M.A.~acknowledges NWO-Rubicon Grant No.~RG86688. V.C.~acknowledges financial 
support provided under the European Union's H2020 ERC Consolidator Grant ``Matter
and strong-field gravity: New frontiers in Einstein's theory", grant agreement no.~MaGRaTh–646597. 
G.K.~acknowledges research support from the National Science Foundation (award no. PHY -- 1701284) 
and Air Force Research Laboratory (agreement no.~10-RI-CRADA-09).

\bibliographystyle{apsrev}
\bibliography{references}

\end{document}